\documentclass[conference]{IEEEtran}
\usepackage{algorithm}
\usepackage{algpseudocode}

\usepackage{blindtext}
\usepackage{multirow}
\usepackage[numbers]{natbib}
\usepackage{notoccite}
\usepackage{graphicx}
\usepackage{subfig}
\usepackage[colorlinks=true]{hyperref}%
\hypersetup{
    linkcolor = blue, citecolor = blue, urlcolor = blue
}

\usepackage{float}

\usepackage{amssymb}
\usepackage{cite}

\ifCLASSINFOpdf
\else
\fi
\usepackage{amsmath}

\usepackage[numbers]{natbib}
\usepackage{xcolor}

\usepackage{url}

\usepackage{amsmath}
\begin{document}
%
\title{Smart Routing with Precise Link Estimation: DSEE-Based Anypath Routing for Reliable Wireless Networking}
%
%
%

\author{\IEEEauthorblockN{Narjes Nourzad and Bhaskar Krishnamachari}

\IEEEauthorblockA{Department of Electrical and Computer Engineering\\
University of Southern California \\
Los Angeles, CA 90089\\
Email:\{nourzad, bkrishna\}@usc.edu}
}

\maketitle

\begin{abstract}

In dynamic and resource-constrained environments, such as multi-hop wireless mesh networks, traditional routing protocols often falter by relying on predetermined paths that prove ineffective in unpredictable link conditions. Shortest Anypath routing offers a solution by adapting routing decisions based on real-time link conditions. However, the effectiveness of such routing is fundamentally dependent on the quality and reliability of the available links, and predicting these variables with certainty is challenging. This paper introduces a novel approach that leverages the Deterministic Sequencing of Exploration and Exploitation (DSEE), a multi-armed bandit algorithm, to address the need for accurate and real-time estimation of link delivery probabilities. This approach augments the reliability and resilience of the Shortest Anypath routing in the face of fluctuating link conditions. By coupling DSEE with Anypath routing, this algorithm continuously learns and ensures accurate delivery probability estimation and selects the most suitable way to efficiently route packets, while maintaining a provable near-logarithmic regret bound. We also theoretically prove that our proposed scheme offers better regret scaling with respect to the network size than the previously proposed Thompson Sampling-based Opportunistic Routing (TSOR).

\end{abstract}

\begin{IEEEkeywords}
Wireless networks, Opportunistic routing, Shortest Anypath routing, Online learning, Multi-armed bandit problem, DSEE.
\end{IEEEkeywords}

%
\IEEEpeerreviewmaketitle

\section{Introduction}
%
%
%
%

Communication networks have become the backbone of modern society, enabling seamless information exchange across a multitude of devices and applications. However, efficient routing of data packets becomes a challenge when faced with the unreliable and unpredictable nature of links in some wireless communication media, such as multi-hop wireless mesh networks. 
The high loss rate and dynamic quality of the links in these wireless networks make traditional routing strategies inadequate. These protocols typically rely on selecting fixed routes based on certain metrics, which can result in packet loss when links fail or degrade unexpectedly. In contrast to these conventional approaches, innovative routing strategies, such as Opportunistic Anypath routing, have emerged \citep{biswas2004opportunistic, shah2005does, liu2009opportunistic, 5061904}.
Anypath routing represents a departure from traditional routing paradigms in that it fully exploits the broadcast nature of wireless communication media. This allows every node to choose several nodes as a set of potential next-hop forwarding nodes basically relaxing the notion of the next hop in traditional wireless routing. When the transmission happens, the best highest priority recipient from the forwarding set opportunistically forwards the message. This is called Anypath routing because each packet from the source may take any of several possible paths to get to the destination. This approach significantly improves the end-to-end throughput, robustness, and cost of data transmission by maximizing the chances of successful packet forwarding at each step, even in the presence of unreliable links~\citep{5061904, Abdollahi2020OpportunisticRM}.
\\

\indent Accurate estimation of link states, particularly data packet delivery probabilities, is crucial for optimizing opportunistic routing in multi-hop wireless mesh networks; however, this task is  challenging due to the dynamic and unpredictable nature of wireless communication environments. Addressing this challenge arises a dilemma between exploring untested links to better estimate their state and exploiting known links for reliable routing, a dilemma central to the multi-armed bandit (MAB) framework~\citep{lattimore2020bandit}. To this end, this work proposes a novel approach based on the Deterministic Sequencing of Exploration and Exploitation (DSEE) algorithm~\citep{Vakili2011DeterministicSO}. DSEE is designed to balance the exploration and exploitation while providing order-optimal \textit{regret}  -- a metric designed to measure the efficiency of a learning algorithm. By incorporating the principles of MABs, our approach ensures that the routing decisions are continually optimized, significantly improving the reliability and efficiency of data packet delivery in such dynamic networks. 
 Moreover, to put this into practice, we employ the Shortest Anypath routing method to efficiently route packets from the source to the destination, leveraging the reliable link state estimations provided by the DSEE algorithm.  
\\

\indent The rest of the paper is organized as follows: Section 2 provides an overview of  the existing literature in the related fields. Section 3 lays the foundational concepts crucial for understanding the DSEE-Anypath method. Section 4 explains our method in detail, introducing the innovative aspects of our approach in anypath routing. Section 5 delves into the theoretical basis of our method, demonstrating its efficiency through regret analysis. Section 6 provides empirical evidence supporting our theoretical findings. Finally, Section 7 concludes the paper, summarizing our findings and suggesting directions for future research. \\ 
 


\section{Related Works}

Multi-armed bandits have been extensively utilized in various domains to explore strategies that optimize decision-making processes. 
The existing body of literature on multi-armed bandits has demonstrated the wide applicability of these strategies in different domains. In the context of dynamic routing through networks, the use of Reinforcement Learning (RL) algorithms, including Multi-Armed Bandit (MAB) approaches, has gained attention due to their capacity to estimate network characteristics and optimize routing decisions in real-time, thereby enhancing the efficiency and robustness of data transmission in such network environments~\citep{9144110, AlRawi2013ApplicationOR}.
\\

\indent The MAB algorithm React-UCB, presented by Santana and Moura~\citep{2023}, is a novel approach designed for optimizing dynamic communication network routing based on the Upper Confidence Bound (UCB) algorithm. The system architecture comprises a controller responsible for gathering network topology data and generating a network representation. Subsequently, this controller transfers the path information to an MAB agent, which selects arms to pull based on piecewise stationary reward distributions in each time step, aiming to minimize end-to-end path delay. The reward obtained from choosing a path is defined as a function of the observed instantaneous path's delay, mappings arms to paths, and relating rewards to path delays.  In contrast, our proposed MAB approach focuses on learning the delivery probability of each link in the network. The ultimate decision regarding path routing is then left to robust routing algorithms such as Shortest Anypath Routing, and the cost of choosing a path is based on the distance each packet traverses. 
\\

\indent The most closely related prior work is that by Huang et al.~\citep{huang2021tsor}, who discuss the limitations of traditional routing in highly dynamic networks and propose an adaptive MAB algorithm called Thompson Sampling-Based Opportunistic Routing (TSOR) for wireless ad hoc networks. In terms of the routing protocol, the authors propose a distributed Bellman-Ford algorithm that is motivated by Thompson sampling. Specifically, the algorithm maintains a probability distribution over the possible link metrics and samples from this distribution to select the next hop for each packet. The algorithm updates the probability distribution based on the observed rewards and routing costs. The authors prove that the expected cumulative regret of the TSOR algorithm is upper-bounded by $O(\log M)$, where $M$ is the number of packets transmitted. However, the derived bounds are cubic with the network size ${N}$ and exponential with the maximum number of neighbor nodes $N_{max}$. 
\\

\indent
To our knowledge, limited studies have succeeded in online learning of link metrics to tackle situations where nodes initially lack knowledge about the network and neighboring nodes. For instance, Bhorkar et al. ~\citep{bhorkar2011adaptive} introduced AdaptOR, a distributed adaptive opportunistic routing mechanism. AdaptOR focuses on minimizing routing costs like energy, delay, and hops. Its core strategy involves each node employing the $\epsilon$-greedy algorithm for next-hop selection, where there's an $\epsilon$ chance of random selection and a 1-$\epsilon$ probability of choosing the node nearest to the destination. While AdaptOR is shown to match the packet-averaged costs of optimal solutions with a large enough volume of transmitted packets, it lacks an analysis of regret, an important metric for assessing convergence speed towards optimal solutions.
\\

\indent
In another study, Liu and Zhao~\citep{6260460} address adaptive shortest-path routing in wireless networks with varying link states. Authors show that by exploiting the structure of the path dependencies, a regret polynomial with network size can be achieved while preserving its optimal logarithmic order with time. However, their approach differs from ours: they treat each path as an arm without observing individual link quality and do not use Shortest Anypath routing,  which ultimately leads to a lower expected cost compared to traditional single-path routing in a wireless network~\citep{5061904}. 
 While there are additional studies delving into learning-based routing, they investigate different scenarios.  For instance, Barve and Kulkarni ~\citep{barve2014multi} implemented reinforcement learning in opportunistic routing for cognitive radio networks. Their approach, however, is based on a different assumption: all nodes in their network scenario share channel availability information through out-of-band beacon messages, which is not the case in our scenario.

\section{Background}
In this section, we delve into the research of {Laufer et al.}~\citep{5061904}, focusing on the Shortest Anypath Routing theory. Additionally, we study the theory of Deterministic Sequencing of Exploration and Exploitation algorithm, the work by {Vakili et al.} \citep{Vakili2011DeterministicSO}, to extend the theory of Shortest Anypath routing to facilitate the learning of link delivery probabilities when such information is unavailable. We borrow most of our notations and concepts from these papers.

\subsection{Single-rate Anypath Routing}
We consider a multi-hop wireless mesh network modeled as a hypergraph $\mathcal{G} = (\mathcal{N}, \mathcal{E})$ where $\mathcal{E}$ is the set of hyperlinks and $\mathcal{N}:= \{1,2,..., N\}$ is the set of nodes. A hyperlink is an ordered pair $(n, J)$ where $n \in \mathcal{N}$ is a node and $J$ is a nonempty subset of $\mathcal{N}$ composed of neighbors of $n$.
\\

The task is to route a sequence of packets $m = 1, 2, ..., M $ from source node \textit{1} to destination node \textit{N} without knowing the complete network topology. Each node $ n \in \mathcal{N}$ transmits the packets in a broadcast way to every node in its forwarding set $J$. The transmission between each node and its forwarding set is characterized by a delivery ratio $p_{nJ}$ and a hyperlink cost $d_{nJ}$.\\

The hyperlink delivery ratio $p_{nJ}$ is defined as the probability that at least one of the nodes in the forwarding set  $J = \{1, 2, ..., N\}$ successfully receives the packet transmitted by node $n$. 
\begin{equation}
    p_{nJ} = 1- P[X_1 = 0, X_2 = 0,..., X_N = 0]
\end{equation}
where $X_j$ is a random variable that follows binomial distribution -- it is equal to 1 if node $j$ successfully receives the packet from node $n$ and 0 otherwise. While assuming that link delivery probabilities are independent is not necessary to demonstrate the optimality of Shortest Anypath Routing, the independence assumption remains a valid approach. This is because the system continuously monitors the channel delivery rate, thereby automatically factoring in any variations in loss rates caused by channel utilization, as outlined in \citep{inproceedings}.\\

The hyperlink cost $d_{nJ}$ defined as $d_{nJ} = 1/p_{nJ}$, represents the average number of transmissions needed to guarantee the correct delivery of a packet to a node within the forwarding set $J$, accounting for a portion of the overall Anypath cost of routing a packet from node $n$ ($D_n = d_{nJ} + D_J $). The remaining anypath cost is defined as a weighted average, taking into account the cost associated with the nodes within the forwarding set. 
\begin{equation}
D_J=\sum_{j \in J} w_{n j} D_j, \text { with } \sum_{j \in J} w_{n j}=1
\end{equation}
The variable $w_{n j}$ indicates the probability that node $j$ operates as a relaying node in cases where node $n$'s packet has been successfully received by at least one node in its forwarding set. This weight is determined by the following expression:

\begin{equation}
w_{n j}=\frac{P[X_1 = 0, X_2 = 0,..., X_{j-1} = 0, X_j = 1]}{1- P[X_1 = 0, X_2 = 0,..., X_N = 0]},
\end{equation}

\noindent with a normalizing constant as a denominator. In the case of independent link delivery probabilities, this expression can be further simplified to:
\begin{equation}
w_{n j}=\frac{p_{n j} \prod_{k=1}^{j-1}\left(p_{n k}\right)}{1-\prod_{j \in J}\left(1-p_{n j}\right)}.
\end{equation}

The Shortest Anypath will always be equivalent to or less costly than the shortest single path. This directly arises from how the anypath scheme is defined: a collection of the paths with the lowest cost among all feasible anypaths between two nodes.
Consequently, when choosing the shortest anypath from all these alternatives, we can confidently state that its expected cost will never exceed that of the shortest single path.

\subsection{Deterministic Sequence of Exploration and Exploitation}
The Multi-Armed Bandit problem represents a category of sequential learning and decision-making with unknown models. In the classic MAB scenario, $N$ distinct arms exist, and during each time step, a single player selects one arm to play, yielding a random reward drawn independently and identically over time from an unknown distribution~\citep{lai1985asymptotically}. The fundamental goal is to establish a sequential arm-selection strategy that optimizes the cumulative expected reward over a given time horizon of length $T$. The main challenge is to reach a balance between exploration and exploitation, where the player must decide whether to opt for an arm that has been explored less but may have potential for future rewards or an arm with a good history of rewards \citep{gittins2011multi, sb}. The performance of a MAB algorithm is evaluated based on its regret, defined as the difference in the total reward it accumulates compared to an optimal decision maker (an oracle) that is aware of the reward probability distributions for all arms.
\\

\indent The key idea behind the Deterministic Sequence of Exploration and Exploitation (DSEE) approach is to separate exploration and exploitation in time, ensuring that learning in the exploration sequence is performed using only reliable observations.  For light-tailed reward distributions, the DSEE achieves the optimal logarithmic order of regret using an exploration sequence with $O(\log T)$ cardinality, given that prior information about the reward models is accessible. However, even in cases where we lack prior knowledge about the exact difference between the true mean of the best and second-best arm, there exists a policy that achieves a regret of  $O(f(T) \log T)$ with high probability by carefully selecting an appropriate sequence $f(t)$ that balances exploration and exploitation.
Logarithmic regret is optimal for any distribution and is achieved by several classic policies, such as UCB1 and Thompson Sampling~\citep{lattimore2020bandit}. However, the DSEE approach has the advantage of being more flexible and applicable to any distribution without knowing the distribution type. \\

\section{DSEE-Anypath Routing}

In our research, we employed a modified variant of the DSEE algorithm to address our specific problem. Unlike the traditional Multi-Armed Bandit framework, where each action corresponds to the selection of a single arm, our approach aligns more closely with the concept explored in combinatorial multi-armed bandits~\citep{gai2012combinatorial}. \\

We leverage the inherent advantage of exploration-exploitation separation within the DSEE framework and enforce specific action selection criteria for each phase. In our methodology, actions taken during the exploitation phase and those during  exploration involve sending (at least) one packet from some nodes or all of the nodes in the network. Notably, the underlying random variables that require observation are exponential in the number of neighbors of each node, corresponding to the joint vector indicating which of the neighboring nodes received a packet sent by a specific node.
More specifically, during the exploration phase, each node transmits a dummy packet to all of its neighboring nodes. Basically, in this phase, we extract samples from the Bernoulli distribution associated with the delivery probability of each link. Once a sufficient amount of exploration determined by  $N \lceil f(t)\text{log}(t) \rceil $ is completed, we transition to the exploitation phase.
The selection of the function $f(t)$ is crucial in our problem formulation since it will heavily influence the delicate balance between exploration and exploitation. The primary constraint on the choice of $f(t)$ is that it must be a positive monotonically increasing function that approaches infinity as $t$ approaches infinity. This constraint provides us with a range of options for selecting $f(t)$. 
It is worth mentioning that even minor inaccuracies in these estimated probabilities can have a notable impact on calculating the anypath cost since these probabilities are inversely related to the weights used to calculate costs for sending packets.
To strike a balance between exploration and exploitation, we choose $f(t)$ as $O(\log{t})$ since our simulations indicate that this provides adequate exploration time for the algorithm. 
\\

In the exploitation phase, we rely on the accurate delivery probability estimates assigned to each link. Initially, we execute a Shortest Anypath First (SAF) algorithm to compute the distance from each node to the destination node and determine the forwarding set for each node. Our routing strategy proceeds by opportunistically forwarding packets and transmitting them through the forwarding set of each node. Different packets may follow different routes based on which nodes receive and forward them at each hop. 
At every node, the packet is broadcasted to all nodes in its forwarding set. If both nodes successfully receive the packet, it is routed through the node with the shortest distance to the destination. In cases where both nodes fail to receive the packet, we resort to re-broadcasting.  The estimation of link delivery probability is updated for all sampled links. The cost of routing each packet is the summation of the distance of each node to the destination. Additionally, we account for additional costs due to packet re-broadcasting when both links fail to deliver the packet effectively. \\

\begin{algorithm}
\caption{DSEE-Anypath Routing}
\begin{algorithmic}[1]
\State \textbf{Initialization}:
\State Set function $f(t)$.
\State  $\mathcal{I}_t \leftarrow \emptyset $.
\While{$t < T$}
\If{$ |\mathcal{I}_t| < N2^{N_{max}} \lceil f(t)\text{log}(t) \rceil$} \Comment{\textit{Exploration Phase}}
\State Add t to the exploration sequence
\For{each node}
    \State  Dummy packet transmission to neighbors.
    \State Link delivery probability update.
\EndFor
\Else
\Comment{\textit{Exploitation Phase}}
\State Compute forwarding sets and distances(SAF).
\For{each packet}
    \State Broadcasting to the current node's forwarding set.
    \If{multiple nodes receive}
        \State Route through shortest distance node.
    \ElsIf{no nodes receive}
        \State Re-broadcast.
        \EndIf
    \State Link delivery probability update.
    \State Add distance to routing cost.
        \If{re-broadcasting}
            \State Add additional costs.
        \EndIf

\EndFor
    \EndIf
\EndWhile
\end{algorithmic}
\end{algorithm}

\section{Performance Analysis}


Inherited from the nature of DSEE, regret is distinctly separated in the exploration and exploitation phases. This separation allows for a detailed analysis of the performance of our method in terms of regret. 
We evaluate our policy based on the path cost for routing packets from source to destination. This approach stems from the cost being shaped by the distance each packet has to traverse through different network nodes. This distance is closely related and influenced by the link delivery probability.
Ultimately, the issue can be traced back to inaccurate estimations of link delivery probabilities. These inaccuracies, in turn, impact the selection of forwarding nodes and, consequently, the chosen routing paths. The outcome is an overall increase in cost and, equivalently, an increase in overall regret.
Regret is quantified as the difference between the cost of the path a genie with complete knowledge of link delivery probabilities would choose at any given time and the cost of the path selected by DSEE-Anypath using its estimated probabilities. 
\\

\indent To achieve optimal network routing, it is crucial to carefully select an exploration sequence to facilitate effective learning and accurate estimation of link delivery probabilities for identifying the best path. At the same time, it should ensure any loss in rewards during exploitation caused by inaccurate path choices have an order no larger than the cardinality of the exploration sequence. \\

\textit{Lemma:} (\textbf{Chernoff-Hoeffding Bound} \citep{doi})  Let $\{X(t)\}_{t=1}^{\infty}$ be i.i.d. random variables drawn from a light-tailed distribution. Let $\overline{X_s}=\left(\sum_{t=1}^s X(t)\right) / s$ and $\theta=\mathbb{E}[X(1)]$. We have, for all $\delta \in\left[0, \zeta u_0\right], a \in\left(0, \frac{1}{2 \zeta}\right]$,
$$
Pr\left(\left|\overline{X_s}-\theta\right| \geq \delta\right) \leq 2 \exp \left(-a \delta^2 s\right)
$$
\\

\textit{Theorem:} (\textbf{DSEE-Anypath routing}) \textit{Let $ \mathcal{I}_t$ be set of times indexes from $1$ to $t-1$ where exploration has already happened. By choosing the cardinality of the exploration to be smaller than $N2^{N_{max}} \lceil f(t)\text{log}(t) \rceil $ at any given $t$, the optimal policy will have regret:}

\begin{equation}
\mathcal{R}_T = O(N2^{N_{max}}f(T)\text{log}(T) \Delta ) 
\end{equation}
\noindent where $N$ is the number of nodes in hypergraph $\mathcal{G}$, $\Delta$ represents the difference between the best available path (path with the lowest cost) and the path selected by the policy, and $N_{max}$ indicating the maximum number of neighbors a node can have, which serves as an upper limit on the size of the forwarding set for each node considers all neighboring nodes as potential candidates for relaying.
The regret bound can be further simplified considering $f(t)=O(\log{t})$, $\Delta = O(N)$:
\begin{equation}
\mathcal{R}_T =O(N^22^{N_{max}}\log^2 T )
\end{equation}

Assuming the wireless links between any two nodes are independent, the exponential number of random variables requiring observation break into a linear form, making our regret bound $\mathcal{R}_T = O(N{N_{max}}f(T)\log T \Delta) = O(N^2{N_{max}}\log^2 T)$.
\\

\textit{Proof:} In the exploration phase, it is easily noticeable that regret increases with the order of the exploration sequence, following a growth pattern of $ N2^{N_{max}} f(T)\log(T) \Delta$ since the regret should grow with the number of random variables governing the problem, in this case, $N2^{N_{max}}$.
\\
Regret in the exploitation phase occurs when the complement of the following event takes place:
\begin{equation}
\mathcal{K}(t) = \left\{\left|\bar{c}_p(t)-c_p\right| \leq \delta, \quad \forall p \in \mathcal{P} \right\}
\end{equation}
in which $\mathcal{P}$ is the set of all anypaths and $c_p$ is the cost of taking an anypath. Notably, careful choice of $\delta$ would result in correct identification of action ranks under $\mathcal{K}$.
\begin{equation*}
\begin{aligned}
\mathcal{R}_T^{exploit} & \leq \Sigma_{t \notin \mathcal{I}_t, t \leq T} \operatorname{Pr}(\overline{\mathcal{K}(t)}) \Delta_p \\
& =\Sigma_{t \notin \mathcal{A}(T), t \leq T} \operatorname{Pr}\left(\exists  p \in \mathcal{P}  \text { s.t. }\left|\bar{c}_p(t)-c_p\right| >\delta\right) \Delta_p \\
& \leq \Sigma_{t \notin \mathcal{A}(T), t \leq T} \Sigma_{p \in \mathcal{P}} \operatorname{Pr}\left(\left|\bar{c}_p(t)-c_p\right|>\delta\right) \Delta_p \\
& \leq 2N2^{N_{max}} \Delta \Sigma_{t \notin \mathcal{A}(T), t \leq T}  \text{exp}(-a  \delta^2 f(t) \log(t))
\end{aligned}
\end{equation*}
\\
From the constraints that are on the choice of $f(t)$:
\begin{equation*}
    \exists \hspace{0.4mm} t_0 \ , \exists\hspace{0.4mm} \beta > 1 \text{ s.t. } \forall t > t_0, \, a\delta^2 f(t) \geq \beta
\end{equation*}
Letting us further simplify the regret bound:
\begin{equation}
\begin{aligned}
\mathcal{R}_T^{exploit} &  \leq 2N2^{N_{max}} \Delta  \big(\sum_{t=1}^{t_0} \exp\left(-a \delta^2 f(t) \log t\right) + \sum_{t=t_0+1}^{\infty} t^{-\beta} \big) \\
& \leq 2N2^{N_{max}} \Delta  \big(t_0 + \frac{1}{\beta - 1} t^{1-\beta}_0 \big)
\end{aligned}
\end{equation}
We applied techniques similar to those outlined in the proof of \textit{Theorem 2} in \citep{Vakili2011DeterministicSO}, and $a$ and $\delta$ are chosen in a way that \textit{Lemma} holds.

The regret bound achieved by the TSOR algorithm in \citep{huang2021tsor} is $\mathcal{R}_M = O(N^34^{N{max}}\log M)$ (assuming $\tau_{max} = 0$ for a centralized scheme). Assuming $M = cNT$ where $M$ is the number of packets and $c \geq 1$, we can write $T  = \frac{M}{cN}$ which is $\Theta(M/N)$ and hence $O(M/N)$ and $\Omega(M/N)$. This enables us to denote TSOR regret bound as $\mathcal{R}_T = O(N^34^{N{max}}\log M) =  O(N^34^{N{max}}(\log N + \log T))$.
When comparing the regret bounds, we can see that our algorithm demonstrates regret bounded by a quadratic function of network size $N$ and exhibits a linear relationship with the parameter $2^{N_{max}}$ in contrast to TSOR with a cubic relation with $N$  and exponential with the parameter $2^{N_{max}}$. This quadratic-linear relationship in our method results in a more favorable scaling behavior for regret in scenarios where the network size grows significantly. However, if we compare in terms of time horizon $T$, for very large time horizons and smaller to medium-sized networks, TSOR's regret bound becomes more favorable. \\

\section{Experimental results}

\begin{figure}[t]
\centering
\includegraphics[width=1.8in]{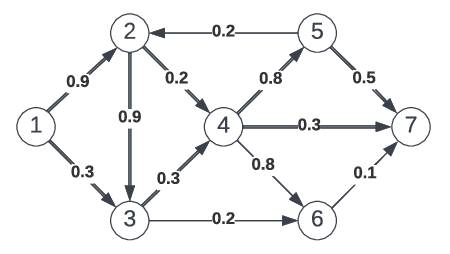}%
\caption{An anypath connecting nodes $1$ and $7$ is shown in double arrows. Every packet sent from  $1$  traverses one of these paths to reach $7$. The numbers on links are delivery success probabilities; in this network, the links are assumed to be independent of each other.}
\label{fig_b1}
\end{figure}

In this section, the results of our experiments are presented with the aim of evaluating the performance of the DSEE-Anypath algorithm in estimating link delivery probabilities for anypath routing in dynamic wireless mesh networks. During the simulations, links were assumed to be independent.
To ensure accurate results, we conducted our experiments by executing the algorithm for a total of 500 epochs. Each epoch consisted of 5000 individual runs in order to closely approximate the true mean of the cost of delivering a packet from source to destination.
To compute the cost, we employed a Shortest Anypath Routing algorithm on the network shown in Figure \ref{fig_b1}. This algorithm was used to identify the optimal paths for routing packets, and it relies on estimated link delivery predictions provided by the DSEE algorithm. \\

\begin{figure}[t]
\centering
\includegraphics[width=3in]{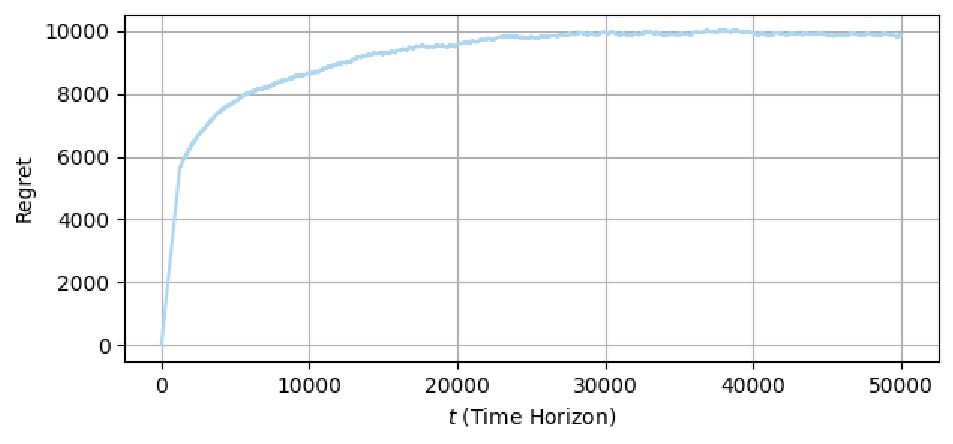}%
\caption{Regret Over Time}
\label{fig_2}
\end{figure}

\begin{figure}[t]
\centering
\includegraphics[width=3in]{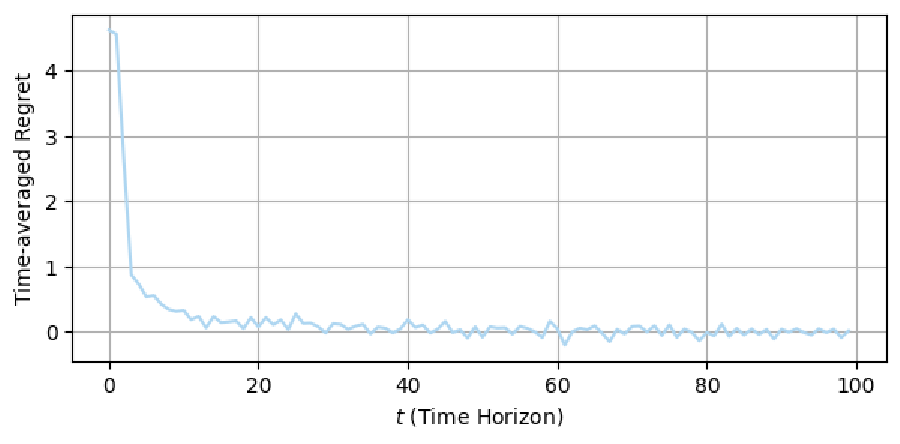}%
\caption{Time-Averaged Regret Over Time}
\label{fig_3}
\end{figure}
The evaluation of our novel algorithm is presented using two metrics that, in the context of multi-armed bandit problems,  are key to understanding  the effectiveness of the provided method. 
Cumulative regret is defined as the total difference in cost between the paths chosen by a genie  and our algorithm over time, and an optimal algorithm should exhibit a logarithmic growth~\citep{lattimore2020bandit}. In contrast, time-averaged regret, the average regret per round, is expected to decrease and converge to zero. This convergence demonstrates the algorithm's improving accuracy in estimating the optimal path as it learns, aligning closer to the genie's knowledge over time. To be more concise, Cumulative regret measures the algorithm's overall performance, while time-averaged regret shows its learning progress and decision-making efficiency over time.
The plot in Figure \ref{fig_2} provides evidence in support of our theorem, illustrating that the cumulative regret exhibits near-logarithmic behavior over time as expected. Moreover, Figure \ref{fig_3} showcases that, on average, our DSEE-Anypath algorithm, without prior link state knowledge, matches the performance of the genie. Additionally, we tested our method in networks with 4 and 10 nodes, observing similar regret trends. However, these results are not presented here due to space constraints.\\

\section{Conclusion and Future Work}
In conclusion, this paper introduced an innovative approach that utilizes the Deterministic Sequencing of Exploration and Exploitation (DSEE) algorithm to address the critical challenge of accurately estimating delivery probabilities for links within Anypath routing, especially in dynamic and resource-constrained multi-hop wireless mesh networks. Throughout the DSEE phase, we use a carefully chosen function, denoted as $f(t)$, to balance the trade-off between exploration and exploitation. Our simulations and empirical findings demonstrate the effectiveness of our approach in terms of the regret bound, thus ensuring an  $ O( N2^{N_{max}} f(T)\log(T) \Delta)$ regret bound. This regret bound, considering $f(t)=O(\log{t})$ and $\Delta = O(N)$, compared to TSOR with regret bound $\mathcal{R}_M = O(N^34^{N{max}}\log M)$ provided in \citep{huang2021tsor}, exhibits a more favorable quadratic-linear scaling with network size $N$ and $2^{N_{max}}$. 
\\

Our future work will focus on incorporating dependent paths, considering that interference can introduce some level of correlation among neighboring links, a common occurrence in high network loads. This will help us provide more accurate estimates of hyperlink delivery ratios by accounting for potential dependencies among the links. Additionally, we plan to evolve this approach into a decentralized scheme that offers increased adaptability and scalability in diverse network environments. Further exploration could involve non-stationary environments, where link estimation during exploration might require techniques like sliding windows or exponentially weighted moving averages.

\section*{Acknowledgment}
This work was supported in part by a seed grant from the USC Office of Research and Innovation. Any views, opinions, and/or findings expressed are those of the author(s) and should not be interpreted as representing the official views or policies of the sponsor.

\ifCLASSOPTIONcaptionsoff
  \newpage
\fi



%



\bibliographystyle{unsrt}
\bibliography{ref}

\begin{thebibliography}{10}

\bibitem{biswas2004opportunistic}
Sanjit Biswas and Robert Morris.
\newblock Opportunistic routing in multi-hop wireless networks.
\newblock {\em ACM SIGCOMM Computer Communication Review}, 34(1):69--74, 2004.

\bibitem{shah2005does}
Rahul~C Shah, Sven Wietholter, Adam Wolisz, and Jan~M Rabaey.
\newblock When does opportunistic routing make sense?
\newblock In {\em Third IEEE International Conference on Pervasive Computing
  and Communications Workshops}, pages 350--356. IEEE, 2005.

\bibitem{liu2009opportunistic}
Haitao Liu, Baoxian Zhang, Hussein~T Mouftah, Xiaojun Shen, and Jian Ma.
\newblock Opportunistic routing for wireless ad hoc and sensor networks:
  Present and future directions.
\newblock {\em IEEE Communications Magazine}, 47(12):103--109, 2009.

\bibitem{5061904}
R.~Laufer, H.~Dubois-Ferriere, and L.~Kleinrock.
\newblock Multirate anypath routing in wireless mesh networks.
\newblock In {\em IEEE INFOCOM 2009}, pages 37--45, 2009.

\bibitem{Abdollahi2020OpportunisticRM}
Mostafa Abdollahi, Farshad Eshghi, Manoochehr Kelarestaghi, and Mozafar
  Bag-Mohammadi.
\newblock Opportunistic routing metrics: A timely one-stop tutorial survey.
\newblock {\em J. Netw. Comput. Appl.}, 171:102802, 2020.

\bibitem{lattimore2020bandit}
Tor Lattimore and Csaba Szepesv{\'a}ri.
\newblock {\em Bandit algorithms}.
\newblock Cambridge University Press, 2020.

\bibitem{Vakili2011DeterministicSO}
Sattar Vakili, Keqin Liu, and Qing Zhao.
\newblock Deterministic sequencing of exploration and exploitation for
  multi-armed bandit problems.
\newblock {\em IEEE Journal of Selected Topics in Signal Processing},
  7:759--767, 2011.

\bibitem{9144110}
Xinyu You, Xuanjie Li, Yuedong Xu, Hui Feng, and Jin Zhao.
\newblock Toward packet routing with fully-distributed multi-agent deep
  reinforcement learning.
\newblock In {\em 2019 International Symposium on Modeling and Optimization in
  Mobile, Ad Hoc, and Wireless Networks (WiOPT)}, pages 1--8, 2019.

\bibitem{AlRawi2013ApplicationOR}
Hasan A.~A. Al-Rawi, Ming~Ann Ng, and Kok lim Alvin~Yau.
\newblock Application of reinforcement learning to routing in distributed
  wireless networks: a review.
\newblock {\em Artificial Intelligence Review}, 43:381 -- 416, 2013.

\bibitem{2023}
Pedro Santana and José Moura.
\newblock A bayesian multi-armed bandit algorithm for dynamic end-to-end
  routing in sdn-based networks with piecewise-stationary rewards.
\newblock {\em Algorithms}, 16(5):233, Apr 2023.

\bibitem{huang2021tsor}
Zhiming Huang, Yifan Xu, and Jianping Pan.
\newblock {TSOR}: Thompson sampling-based opportunistic routing.
\newblock {\em IEEE Transactions on Wireless Communications},
  20(11):7272--7285, 2021.

\bibitem{bhorkar2011adaptive}
Abhijeet~A Bhorkar, Mohammad Naghshvar, Tara Javidi, and Bhaskar~D Rao.
\newblock Adaptive opportunistic routing for wireless ad hoc networks.
\newblock {\em IEEE/ACM Transactions On Networking}, 20(1):243--256, 2011.

\bibitem{6260460}
Keqin Liu and Qing Zhao.
\newblock Adaptive shortest-path routing under unknown and stochastically
  varying link states.
\newblock In {\em 2012 10th International Symposium on Modeling and
  Optimization in Mobile, Ad Hoc and Wireless Networks (WiOpt)}, pages
  232--237, 2012.

\bibitem{barve2014multi}
Sunita~S Barve and Parag Kulkarni.
\newblock Multi-agent reinforcement learning based opportunistic routing and
  channel assignment for mobile cognitive radio ad hoc network.
\newblock {\em Mobile Networks and Applications}, 19:720--730, 2014.

\bibitem{inproceedings}
Richard Draves, Jitendra Padhye, and Brian Zill.
\newblock Routing in multi-radio, multi-hop wireless mesh networks.
\newblock pages 114--128, 09 2004.

\bibitem{lai1985asymptotically}
Tze~Leung Lai and Herbert Robbins.
\newblock Asymptotically efficient adaptive allocation rules.
\newblock {\em Advances in applied mathematics}, 6(1):4--22, 1985.

\bibitem{gittins2011multi}
John Gittins, Kevin Glazebrook, and Richard Weber.
\newblock {\em Multi-armed bandit allocation indices}.
\newblock John Wiley \& Sons, 2011.

\bibitem{sb}
Richard~S Sutton and Andrew~G Barto.
\newblock {\em Reinforcement learning: An introduction}.
\newblock MIT press, 2018.

\bibitem{gai2012combinatorial}
Yi~Gai, Bhaskar Krishnamachari, and Rahul Jain.
\newblock Combinatorial network optimization with unknown variables:
  Multi-armed bandits with linear rewards and individual observations.
\newblock {\em IEEE/ACM Transactions on Networking}, 20(5):1466--1478, 2012.

\bibitem{doi}
Rajeev Agrawal.
\newblock The continuum-armed bandit problem.
\newblock {\em SIAM Journal on Control and Optimization}, 33(6):1926--1951,
  1995.

\end{thebibliography}

%








\end{document}